\documentclass[a4paper]{jpconf}
\usepackage{graphicx}
\usepackage{amsmath,amssymb,amsfonts}
\usepackage{subfigure}
\usepackage{listings}

\newcommand{\GOLEM}{{\textsc{Go\-Sam}}}
\newcommand{\GOSAM}{{\textsc{Go\-Sam}}}
\newcommand{\GOLEMVC}{{\texttt{go\-lem95C}}}
\newcommand{\QGRAF}{{\texttt{QGRAF}}}
\newcommand{\FORM}{{\texttt{FORM}}}
\newcommand{\SPINNEY}{{\texttt{spin\-ney}}}
\newcommand{\HAGGIES}{{\texttt{hag\-gies}}}
\newcommand{\SAMURAI}{{\textsc{Sa\-mu\-rai}}}
\newcommand{\PYTHON}{{\texttt{Py\-thon}}}
\newcommand{\bea}{\begin{eqnarray}}
\newcommand{\eea}{\end{eqnarray}\noindent}
\newcommand{\nn}{\nonumber}
\newcommand{\bcen}{\begin{center}}
\newcommand{\ecen}{\end{center}}
\def\url#1{\texttt{#1}}

\def\eps{\epsilon}

\begin{document}
\title{GoSam: A program for automated one-loop Calculations}

\author{G.~Cullen}
\address{Deutsches Elektronen-Synchrotron DESY, Zeuthen, Germany}

\author{N.~Greiner, G.~Heinrich\footnote{Speaker; Talk given at the International Workshop on Advanced Computing and
Analysis Techniques in Physics Research (ACAT), Uxbridge, London, September 2011.},  T.~Reiter}
\address{Max-Planck-Institute for Physics,  Munich, Germany}

\author{G.~Luisoni}
\address{Institute for Particle Physics Phenomenology, University of Durham, UK}

\author{P.~Mastrolia}
\address{Max-Planck Institute for Physics, Munich, Germany;\\
Dipartimento di Fisica, Universita di Padova, Italy}

\author{G.~Ossola}
\address{New York City College of Technology, City University of New York}

\author{F.~Tramontano}
\address{CERN, Geneva, Switzerland}


\begin{abstract}
The program package \GOSAM{} is presented which aims at the automated calculation of 
one-loop amplitudes for multi-particle processes.
The amplitudes are generated in terms of Feynman diagrams and 
can be reduced using either D-dimensional integrand-level decomposition
or tensor reduction, or a combination of both. 
\GOSAM{} can be used to calculate one-loop corrections to both QCD and electroweak theory, 
and model files for theories Beyond the Standard Model can be linked as well.
A standard interface to programs calculating real radiation is also included. 
The flexibility of the program is demonstrated by various examples. 
\end{abstract}

\section{Introduction}

Precise theory predictions play an important role in the analysis and
the interpretation of collider physics data, in the search for new particles 
as well as to constrain model parameters at a later stage.
Therefore it is desirable to have predictions at next-to-leading order (NLO)
in perturbation theory as a standard at the LHC.
However, this can only be achieved if a level of automation for NLO 
predictions is reached 
which comes close to the one we have for leading order tools.

The need for an automation of NLO calculations has been noticed some time ago
and lead to public programs like FeynArts\,\cite{Hahn:2000kx} and
 QGraf\,\cite{Nogueira:1991ex} 
for diagram generation and FormCalc/LoopTools\,\cite{Hahn:1998yk} and 
{\small GRACE} \,\cite{Belanger:2003sd} for 
the automated calculation of NLO corrections. 
However, calculations of one-loop amplitudes with more than four external legs  
were still tedious case-by-case calculations. Only very recently, conceptual 
and technical advances in multi-leg one-loop calculations 
opened the door to the possibility of an {\it automated} generation 
and evaluation of multi-leg one-loop amplitudes.
As a consequence,  public NLO tools containing
a collection of hard-coded individual processes, like  e.g. 
MCFM\,\cite{Campbell:1999ah,Campbell:2011bn},
VBFNLO\,\cite{Arnold:2008rz,Arnold:2011wj,Campanario:2011cs}, 
MC@NLO\,\cite{Frixione:2002ik,Frixione:2010wd}, POWHEG-\hspace{0pt}Box \cite{Frixione:2007vw,Alioli:2010xd}, POWHEL\,\cite{Kardos:2011qa,Garzelli:2011vp,Kardos:2011na},
are now being complemented by flexible automated tools\,\cite{Cullen:2011ac,vanHameren:2009dr,Hirschi:2011pa,
Bevilacqua:2011xh,HahnAcat} 
which have as their aim that basically any process which may turn out to be important for 
the comparison of LHC findings to theory can be evaluated at NLO accuracy.

In this talk, we present the program package \GOSAM{}\cite{Cullen:2011ac} 
which allows the automated calculation of 
one-loop amplitudes for multi-particle processes. 
Amplitudes are expressed in terms of Feynman diagrams, where the integrand is generated analytically using \QGRAF~\cite{Nogueira:1991ex}, \FORM~\cite{Vermaseren:2000nd},
\SPINNEY~\cite{Cullen:2010jv} and \HAGGIES~\cite{Reiter:2009ts}. 
The individual program tasks are steered via python scripts, while the user only needs to edit an ``input card" to specify the details of the process to be calculated,
and launch the generation of the source code and its compilation, without having to worry about internal details of the 
code.

The program offers the option to use different reduction techniques:
either the unitarity-based integrand reduction~\cite{Ossola:2006us,Ellis:2008ir}  
as implemented in \SAMURAI~\cite{Mastrolia:2010nb} 
or traditional tensor reduction as implemented in
{\tt golem95C}~\cite{Binoth:2008uq,Cullen:2011kv} interfaced through
tensorial reconstruction at the integrand level~\cite{Heinrich:2010ax},
or a combination of both.
It can be used to calculate one-loop corrections within both QCD and electroweak theory. 
Beyond the Standard Model theories can be interfaced using
FeynRules~\cite{Degrande:2011ua} or \texttt{LanHEP}~\cite{Semenov:2010qt}.
The  Binoth Les Houches interface\,\cite{Binoth:2010xt} to programs providing the real radiation 
contributions is also included. 

\section{Theoretical framework}

\subsection{Generation of the diagrams}
For the diagram generation 
we use the program \QGRAF~\cite{Nogueira:1991ex} and 
supplement it by adding another level of analysing and filtering the diagrams,
written in \PYTHON{}. 
This allows for example to drop diagrams whose colour factor turns out to be zero, 
or to determine the signs for diagrams with Majorana fermions.

Within our framework, \QGRAF{} generates three sets
of output files: an expression for each diagram to be processed with
\FORM~\cite{Vermaseren:2000nd}, \PYTHON{} code to draw all diagrams,
and \PYTHON{} code to compute the properties of each diagram. 
The information about the model
is either read from the built-in Standard Model file or
is generated from a user defined \texttt{LanHEP}~\cite{Semenov:2010qt}
or Universal FeynRules Output (\texttt{UFO})~\cite{Degrande:2011ua} file.

The program also produces a  \LaTeX{} file which contains graphical
representations of all diagrams as well as a summary of the helicity and colour basis used.

\subsection{Code generation }

The amplitude is generated in terms of algebraic expressions based on 
Feynman diagrams and then processed with a \FORM{} program, using {\tt spinney}~\cite{Cullen:2010jv}
for the spinor algebra. 
In \GOLEM{} we have implemented the 't~Hooft-\hspace{0pt}Veltman
scheme (HV) and
dimensional reduction~(DRED). In both schemes all external vectors
(momenta and polarisation vectors) are kept in four dimensions, while 
internal vectors are kept in the $n$-dimensional vector space ($n=4-2\eps$).
We adopt the conventions used in~\cite{Cullen:2010jv}, where
$\hat{k}$ denotes the four-dimensional projection of an 
$n$-dimensional vector $k$. The $(n-4)$-dimensional orthogonal projection
is denoted by~$\tilde{k}$. For the loop momentum $q$ we introduce
the symbol $\mu^2=-\tilde{q}^2$, such that
\begin{equation}
q^2=\hat{q}^2+\tilde{q}^2=\hat{q}^2-\mu^2.
\end{equation}

To prepare the numerator functions of the one-loop diagrams for their numerical evaluation,
we separate the symbol $\mu^2$ and dot products involving the
momentum $\hat{q}$ from all other factors. All subexpressions which do
not depend on either $\hat{q}$ or $\mu^2$ are substituted by abbreviations,
which are evaluated only once per phase space point.
Each of the two parts is then processed by \HAGGIES~\cite{Reiter:2009ts},
which generates optimised \texttt{Fortran95} code for the numerical evaluation.
For each diagram we generate an interface to
\SAMURAI~\cite{Mastrolia:2010nb}, \GOLEMVC~\cite{Cullen:2011kv} and/or
\texttt{PJFRY}~\cite{Fleischer:2010sq}. 

Our standard choice for the reduction is to use \SAMURAI{}~\cite{Mastrolia:2010nb}, 
which provides
a very fast and stable reduction in a large part of the phase space.
Furthermore, \SAMURAI{} reports to the client code if the quality
of the reconstruction of the numerator suffices the numerical requirements. 
In \GOLEM{} we use this
information to trigger an alternative reduction with either
\GOLEMVC~\cite{Cullen:2011kv} or \texttt{PJFRY}~\cite{Fleischer:2010sq} whenever
these reconstruction tests fail. 
This combination of on-shell techniques and traditional tensor reduction
is achieved using tensorial reconstruction at the integrand
level~\cite{Heinrich:2010ax}. 
The tensorial reconstruction not only can cure numerical instabilities, but in some cases also
can reduce the computational cost of the reduction.
Since the reconstructed numerator is typically of a form where 
kinematics and loop momentum dependence are already separated, 
the use of a reconstructed numerator tends to be faster than the original procedure, 
in particular in cases with a large number of legs and low rank. 


\subsection{Rational terms }

Terms containing the symbols $\mu^2$ or $\varepsilon$ 
in the numerator of the integrands 
can lead to a so-called
$R_2$ term~\cite{Ossola:2008xq}, which contributes to the rational part 
of the amplitude.
We generate the $R_2$ part along with all other
contributions without the need to seperate the different parts.
In addition, we provide an \emph{implicit} and
an \emph{explicit} construction of the $R_2$ terms, using the fact that 
there are two ways of splitting the numerator function:
\bea
\label{eq:r2:a}
\mathcal{N}(\hat{q},\mu^2,\varepsilon)&=&
\mathcal{N}_0(\hat{q},\mu^2)+
\varepsilon\mathcal{N}_1(\hat{q},\mu^2)+\varepsilon^2\mathcal{N}_2(\hat{q},\mu^2)\\
&&\mbox{or, alternatively,}\nn\\
\label{eq:r2:b}
\mathcal{N}(\hat{q},\mu^2,\varepsilon)&=&
\hat{\mathcal{N}}(\hat{q})+
\tilde{\mathcal{N}}(\hat{q},\mu^2,\varepsilon).
\eea
The implicit
construction uses the splitting of Eq.~(\ref{eq:r2:a}) and treats
all three numerator functions~$\mathcal{N}_i$ on equal grounds.
Each of the three terms is reduced seperately in a numerical reduction
and the Laurent series of the three results are added up taking into
account the powers of $\varepsilon$. 

The explicit construction of $R_2$
is based on the assumption that each term in
$\tilde{\mathcal{N}}$ in Eq.~(\ref{eq:r2:b}) contains at least one
power of $\mu^2$ or $\varepsilon$. The expressions for those integrals
are relatively simple and known explicitly. Hence, the part of the amplitude
which originates from $\tilde{\mathcal{N}}$ is computed analytically whereas
the purely four-\hspace{0pt}dimensional part $\hat{\mathcal{N}}$ is passed to the numerical
reduction.

In the program, possible options for $R_2$ are \texttt{r2=implicit},\texttt{explicit}, \texttt{off} 
and \texttt{only}.
Using \texttt{r2=only} discards everything but
the $R_2$ term 
and puts \GOLEM{} in the position of providing $R_2$ terms
to supplement other codes which work entirely in four dimensions.

\subsection{Conventions}

To be specific, we consider the
case where the user wants to compute QCD corrections. 
In the case of electroweak corrections, 
the analogous conventions apply except that the strong coupling $g_s$ is replaced by $e$.
In the QCD case, the tree-level
matrix element squared can be written as
\begin{equation}\label{eq:amp0:def}
\vert\mathcal{M}\vert_{\text{tree}}^2=\mathcal{A}_0^\dagger\mathcal{A}_0=
(g_s)^{2b}\cdot a_0\;, 
\end{equation}
where $b=0$ is also possible.
The matrix element at one-loop level, i.e. the
interference term between tree-level and one-loop, can be written as
\begin{multline}\label{eq:amp1:def}
\vert\mathcal{M}\vert_{\text{1-loop}}^2=
\mathcal{A}_1^\dagger\mathcal{A}_0+
\mathcal{A}_0^\dagger\mathcal{A}_1=
2\cdot\Re(\mathcal{A}_0^\dagger\mathcal{A}_1)=
\frac{\alpha_s(\mu)}{2\pi}\frac{(4\pi)^\varepsilon}{\Gamma(1-\varepsilon)}
\, (g_s)^{2b}\,\left[%
c_0+\frac{c_{-1}}{\varepsilon}+\frac{c_{-2}}{\varepsilon^2}
+\mathcal{O}(\varepsilon)%
\right]\;.
\end{multline}
A call to the subroutine \texttt{samplitude} returns an array
consisting of the four numbers $(a_0, c_0, c_{-1}, c_{-2})$
in this order. 
The average over initial state colours and helicities is included 
in the default setup. Renormalisation is included depending on the options 
chosen by the user, for a more detailed description we refer to \cite{Cullen:2011ac}.
In cases where the process is loop induced, i.e. the tree level amplitude is absent, 
the program returns the values for $\mathcal{A}_1^\dagger\mathcal{A}_1$ 
 $$\left(\frac{\alpha_s(\mu)}{2\pi}\frac{(4\pi)^\varepsilon}{\Gamma(1-\varepsilon)}\right)^2$$ 
has been extracted.

\section{Installation and Usage}

\subsection{Installation}
The user can download the code \GOLEM{} either as a tar-ball
    or from the subversion repository
    at
    \bcen
    \url{http://projects.hepforge.org/gosam/}\, .
    \ecen
    The build process and in\-stal\-la\-tion of \GOLEM{} is controlled by
    \PYTHON{} \texttt{Dist\-utils}, while the build process for the libraries 
    \SAMURAI{} and \GOLEMVC{}
    is controlled by \texttt{Autotools}.
To install \GOLEM{}, the user needs to run
\begin{lstlisting}
python setup.py install --prefix MYPATH
\end{lstlisting}
If \texttt{MYPATH} is different from the system default, 
the environment
variables \texttt{PATH}, \texttt{LD\_LIB\-RA\-RY\_PATH} and \texttt{PYTHONPATH}
might have to be set accordingly. For more details we direct the user to
the \GOLEM{} reference manual coming with the code.
    
The program is designed to run in any modern Linux/Unix environment;
we expect that \PYTHON~($\geq2.6$), Java~($\geq1.5$) and Make are installed
on the system. Furthermore, a {\texttt Fortran}\,95 compiler is required in order to
compile the generated code. 

On top of a standard Linux environment, the programs
\FORM~\cite{Vermaseren:2000nd}, 
version~$\geq3.3$, and
\QGRAF~\cite{Nogueira:1991ex} need to be installed on the system.
Further, at least one of the libraries
\SAMURAI{}~\cite{Mastrolia:2010nb} and \GOLEMVC{}~\cite{Cullen:2011kv}
needs to be present at compile time of the generated code.    
For the user's convenience we have prepared a package containing
    \SAMURAI{} and \GOLEMVC{} together with the integral libraries
    \texttt{One\-LOop}\,\cite{vanHameren:2010cp},
    \texttt{QCD\-Loop}\,\cite{Ellis:2007qk} and
    \texttt{FF}\,\cite{vanOldenborgh:1989wn}.
   The package \texttt{gosam-\hspace{0pt}contrib-\hspace{0pt}1.0.tar.gz}
   containing all these libraries is available
   for download from \url{http://projects.hepforge.org/gosam/}.

\subsection{Usage}

In order to generate the code for a process,
the user needs to prepare an input file, 
called \textit{process card} in the following, which contains
\begin{itemize}
\item process specific information, such as a list of initial and
      final state particles, their helicities (optional) 
      and the order of the coupling constants;
\item scheme specific information and approximations, such as
      the regularisation and renormalisation schemes,
      the underlying model,
      masses and widths which are set to zero, 
      the selection of subsets of diagrams, etc; 
\item system specific information, such as paths to programs and libraries
      or compiler options;
\item optional information for optimisations which control the code generation.
\end{itemize}
For example, to compute the QCD corrections for the
process $u\bar{u}\to gZ^0\to g\,e^-e^+$, the first part of the process card
looks as shown in Table\,\ref{Tab:process}:
\begin{table}
\caption{Part of the process card defining the initial and final state and the type and 
order of the coupling at LO and NLO\label{Tab:process}}
\begin{tabular}{ll}
process\_path=&qqgz\\
in=&u,u$\sim$\\
out=&g,\,e-,\,e+\\
helicities=&$+\,-\,+\,-\,+,\,+\,-\,-\,-\,+,\,-\,+\,+\,-\,+,\,-\,+\,-\,-\,+$\\
order=&QCD,\,1,\,3
\end{tabular}
\end{table}
The first line defines the (relative) path to the directory where the
process files will be generated. \GOLEM{} expects that this directory
has already been created.
Lines 2 and~3 define the initial and final state of the process in
terms of field names, which are defined in the model file. Besides
the field names one can also use PDG codes~\cite{Yost:1988ke,Caso:1998tx}
instead. 
For more details we refer to \cite{Cullen:2011ac} and the reference manual.

\subsection{Code generation}
If the process card is called \texttt{gosam.in}, it can be invoked by 
\texttt{gosam.py}: 
\begin{lstlisting}
mkdir qqgz
gosam.py gosam.in
cd qqgz
\end{lstlisting}
All further steps are controlled by the generated make files;
in order to generate and compile all files relevant for the matrix element
one needs to invoke
\begin{lstlisting}
make compile
\end{lstlisting}
The generated code can be tested with the program \texttt{matrix/test.f90}.
The following sequence of commands will compile and run the test program:
\begin{lstlisting}
cd matrix
make test.exe
./test.exe
\end{lstlisting}
The last lines of the program output should look as follows.
\begin{lstlisting}
#             LO:  0.3450350717601E-06
# NLO, finite part  -10.77604823456547
# NLO, single pole  -19.98478948141949
# NLO, double pole  -5.666666665861926
# IR,  single pole  -19.98478948439310
# IR,  double pole  -5.666666666666666
\end{lstlisting}
The printed numbers are, in this order, $a_0$, $c_0/a_0$, $c_{-1}/a_0$,
$c_{-2}/a_0$ and the pole parts calculated from the infrared insertion
operator~\cite{Catani:1996vz,Catani:2000ef}.
One can also generate a pictorial representation of all generated diagrams using
the command
\begin{lstlisting}
make doc
\end{lstlisting}
which generates the file \texttt{doc/process.ps}

\subsection{Interfacing the code}
\subsubsection{Using the BLHA Interface}

The so-called \emph{Binoth Les Houches Accord} (BLHA)~\cite{Binoth:2010xt}
defines an interface for a standardized communication between
one-loop programs (OLP) and Monte Carlo (MC) tools. 
\GOLEM{} can act as an OLP in the framework of the BLHA, such that 
the calculation of complete cross sections is straightforward.

In general, the MC writes an order file, called for  example 
\texttt{olp\_order.lh}, and invokes the script \texttt{gosam.py}
as follows:
\begin{lstlisting}
gosam.py --olp olp_order.lh
\end{lstlisting}
The invocation of \texttt{gosam.py} generates a set of files which
can be compiled with a generated make file. The BLHA routines
are defined in the \texttt{Fortran} module \texttt{olp\_module} but can also
be accessed from \texttt{C} programs. 

\subsubsection{Using External Model Files}

\GOLEM{} can also make use of model files generated
by either \texttt{Feynrules}~\cite{Christensen:2008py}
in the \texttt{UFO} format~\cite{Degrande:2011ua} or by
{\texttt LanHEP}~\cite{Semenov:2010qt}. 
The particles can be specified by their PDG code.
Details about how to import thes model files are described in
the \GOLEM{} reference manual. 
Precompiled \texttt{MSSM\_UFO} and \texttt{MSSM\_LHEP} files 
can also be found in the subdirectory \texttt{examples/model}.
The \texttt{examples} directory also contains examples where 
\texttt{UFO} or {\texttt LanHEP} model files are imported.

\section{Examples}
The code generated by \GOLEM{} has been compared to a considerable number of processes 
available in the literature, as listed in Table \ref{Table:compare}. 
Many of these processes are also included as examples in the code, 
including reference values. 

\begin{figure}[htb]
\subfigure[Transverse momentum of the leading jet for $W^{-}$+\,jet production at LHC with $\sqrt{s}=7$~TeV.]{\label{fig:gosamsherpa:pt}%
\includegraphics[width=0.53\textwidth]{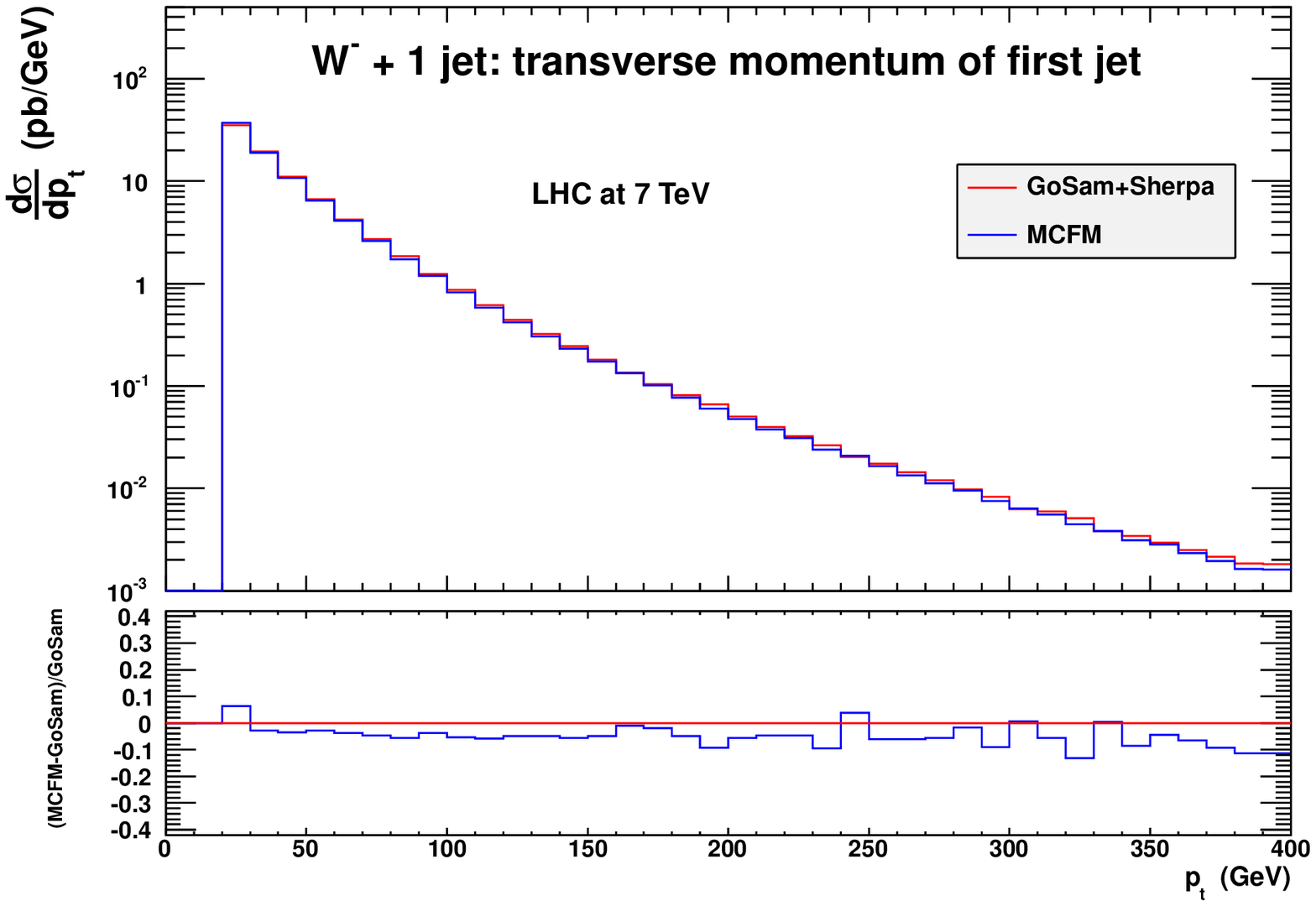}}
\subfigure[Pseudorapidity of the leading jet for $W^{-}$+\,jet production
at LHC with $\sqrt{s}=7$~TeV.]{\label{fig:gosamsherpa:eta}%
\includegraphics[width=0.53\textwidth]{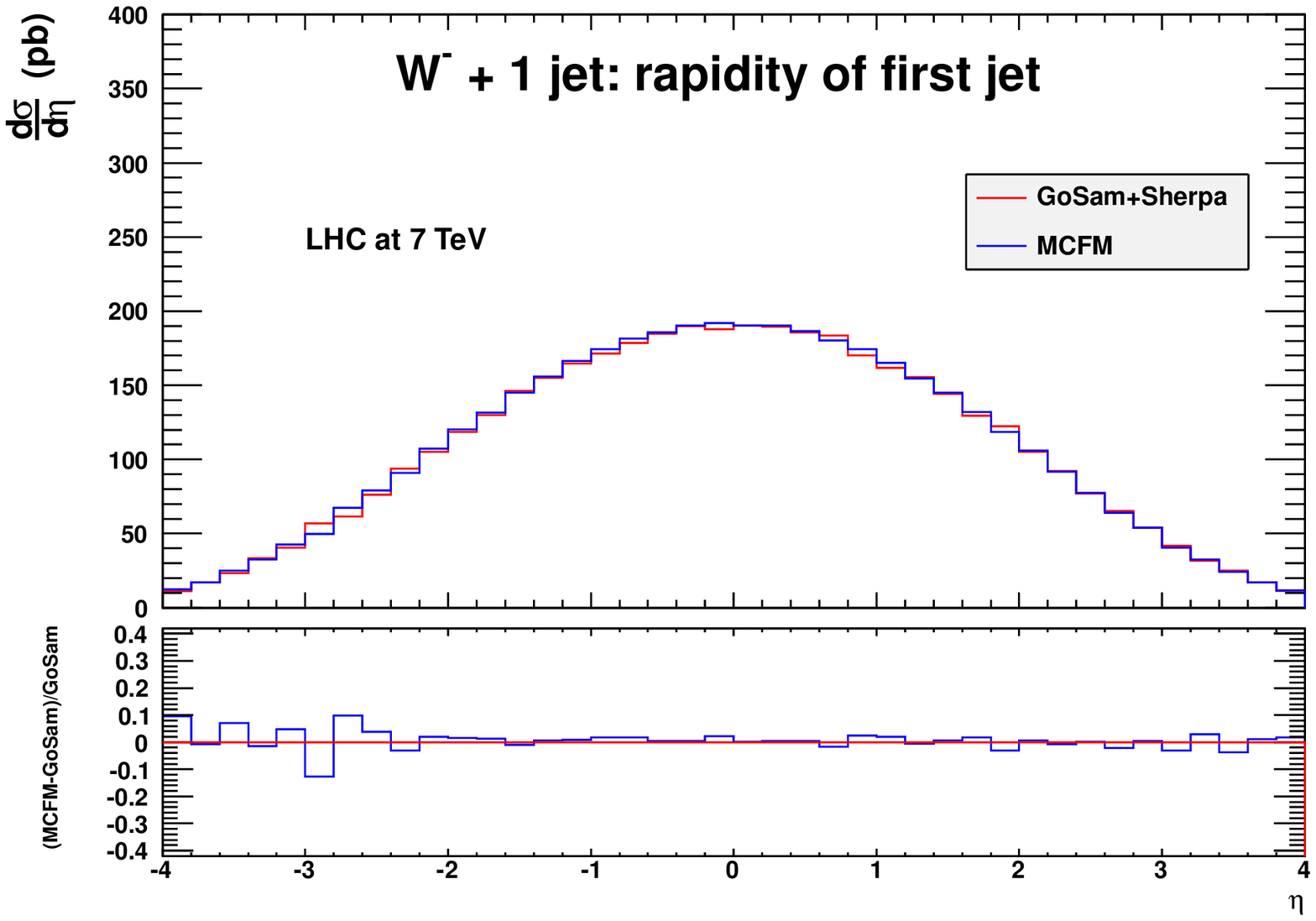}}
\caption{NLO calculation of $W^{-}+$\,jet production at LHC using \GOLEM{} 
interfaced with \texttt{SHERPA} via the Binoth Les Houches interface and compared to MCFM.}
\end{figure} 

\begin{table}
\begin{center}
\begin{tabular}{|ll|}
\hline
\bf process&\bf checked with Ref.\\
\hline
$e^+e^-\to u\overline{u}$&\cite{Ellis:1991qj}\\
$e^+e^-\to t\overline{t}$&\cite{Jersak:1981sp,Catani:2002hc}, own analytic calculation\\
$u\overline{u}\to d\overline{d}$&\cite{Ellis:1985er,Hirschi:2011pa}\\
$g g \to gg$&\cite{Binoth:2006hk}\\
$g g \to gZ$&\cite{vanderBij:1988ac}\\
$d\overline{d}\to t\overline{t}$&\cite{Hirschi:2011pa}, MCFM \cite{Campbell:1999ah,Campbell:2000bg} \\
$g g \to t\overline{t}$&\cite{Hirschi:2011pa}, MCFM \cite{Campbell:1999ah,Campbell:2000bg}\\
$b g \to H\,b$&\cite{Campbell:2002zm,Hirschi:2011pa}\\
$\gamma \gamma \to \gamma \gamma $&\cite{Gounaris:1999gh}\\
$u\overline{d}\to e^-\overline{\nu}_e$&\cite{Hirschi:2011pa}\\
$u\overline{d}\to e^-\overline{\nu}_e\,g$&\cite{Hirschi:2011pa}\\
$e^+e^-\to e^+e^-\gamma$ (QED)&\cite{Actis:2009uq}\\
$pp \to H\,t\overline{t}$&\cite{Hirschi:2011pa}\\
$pp\to W^+W^+jj$&\cite[v3]{Melia:2010bm}\\
$pp\to W^\pm\,j$ (QCD corr.)&MCFM \cite{Campbell:1999ah,Campbell:2000bg}\\
$pp\to W^\pm\,j$ (EW corr.)& for IR poles: Eqs.~(67),(70) of~\cite{Kuhn:2007cv}, \cite{Gehrmann:2010ry}\\
$pp\to b\overline{b} b\overline{b}$ &\cite{Binoth:2009rv,Greiner:2011mp}\\
$pp\to W^+W^- b\overline{b}$ &\cite{vanHameren:2009dr,Hirschi:2011pa}\\
$u\overline{u} \to t\overline{t}b\overline{b}$&\cite{vanHameren:2009dr,Hirschi:2011pa}\\
$gg \to t\overline{t}b\overline{b}$&\cite{vanHameren:2009dr,Hirschi:2011pa}\\
$u\overline{d} \to W^+ ggg$&\cite{vanHameren:2009dr}\\
\hline
\end{tabular}
\end{center}
\caption{Processes for which \GOLEM{} has been compared to the literature.\label{Table:compare}}
\end{table}

As an example for te  {\it Binoth Les Houches Accord} interface of \GOSAM{} we present results for 
the QCD corrections to $W^{-}+1$ jet production, obtained by linking \GOSAM{} with 
\texttt{SHERPA}~\cite{Gleisberg:2008ta}. 
Furthermore, \texttt{SHERPA} offers the possibility to match NLO calculations with a 
parton shower~\cite{Hoche:2010pf,Hoeche:2011fd}. 
Results for the transverse momentum and rapidity distribution 
of the leading jet are shown in Figs.~\ref{fig:gosamsherpa:pt} and~\ref{fig:gosamsherpa:eta}.
The comparison with MCFM\,\cite{Campbell:1999ah,Campbell:2011bn} shows perfect agreement.


\section{Conclusions}

We have presented the program \GOLEM{} 
which can produce  code  for
the evaluation of one-loop matrix elements for multi-particle processes 
in an automated way.
The program is publicly available at
\url{http://projects.hepforge.org/gosam/} 
and can be used to calculate one-loop amplitudes 
within QCD, electroweak theory, or other models 
which can be imported  via an interface to
 LanHEP  or FeynRules. 
 Monte Carlo programs for the real radiation can be easily linked 
 via the interface defined by the {\it Binoth Les Houches Accord}.
 
The amplitudes are generated in terms of Feynman diagrams
and can be reduced by unitarity based reduction
at integrand level or traditional tensor reduction, or a combination of the two 
approaches. This feature makes the program extremely robust against numerical instabilities.
The user can choose among different libraries for the master integrals, and 
the setup is such that  other libraries can be linked easily.

The calculation of the rational terms can proceed either 
together with the same numerical reduction as the rest of the amplitude, 
or  before any reduction,  
using analytic information on the integrals which can potentially give rise to 
a rational part.
Moreover, the \GOSAM{} generator can  produce code for processes
which include unstable particles, i.e. intermediate states with complex masses.

 \GOSAM{} is very well suited for the automated matching of Monte Carlo programs 
 to NLO virtual amplitudes, and therefore
 can be used as a module to produce 
 differential cross sections for multi-particle processes 
 which can be compared directly to experiment. 
 Thus \GOSAM{} can contribute to the goal of automating NLO corrections 
 to an extent where NLO tools become a standard  for data analysis
 at the LHC.

\subsection*{Acknowledgments}
G.C. and G.L. were supported by the British Science and Technology Facilities
Council (STFC).
The work of G.C. was supported by DFG
Sonderforschungsbereich Transregio 9, Computergest\"utzte Theoretische Teilchenphysik.
N.G. was supported in part by the U.S. Department of Energy under contract
No. DE-FG02-91ER40677.
P.M. and T.R. were supported by the Alexander von
Humboldt Foundation, in the framework of the Sofja Kovaleskaja Award Project
``Advanced Mathematical Methods for Particle Physics'', endowed by the German
Federal Ministry of Education and Research.
The work of G.O. was supported in part by the National Science Foundation
under Grant PHY-0855489 and PHY-1068550.
The research of F.T. is supported by Marie-Curie-IEF, project:
``SAMURAI-Apps''.
We also acknowledge the support of the Research Executive Agency (REA)
of the European Union under the Grant Agreement number
PITN-GA-2010-264564 (LHCPhenoNet).


\section*{References}

\providecommand{\newblock}{}

\end{document}